# Light emission properties of mechanical exfoliation induced extended defects in hexagonal boron nitride flakes.

G. Ciampalini[1,2,3]*, C. V. Blaga[4]*. N. Tappy[4], S. Pezzini[3], Watanabe[5], Taniguchi[6], F Bianco[3], S. Roddaro[1,3], A. Fontcuberta i Morral[4,7], F. Fabbri[3]

* These authors contributed equally to the manuscript.

[1] Dipartimento di Fisica E. Fermi, Università di Pisa, Largo B. Pontecorvo 3, I-56127 Pisa, Italy

[2] Graphene Labs, Istituto Italiano di Tecnologia, Via Morego 30, I-16163 Genova, Italy

[3] NEST, Istituto Nanoscienze – CNR, Scuola Normale Superiore, Piazza San Silvestro 12, 56127 Pisa, Italy

[4] Laboratoire des Matériaux Semiconducteurs, Institute of Materials, Faculty of Engineering, École Polytechnique Fédérale de Lausanne, 1015 Lausanne, Switzerland

[5] Research Center for Functional Materials, National Institute for Materials Science, 1-1 Namiki, Tsukuba, 305-0044, Japan

[6] International Center for Materials Nanoarchitectonics, National Institute for Materials Science, 1-1 Namiki, Tsukuba 305-0044, Japan

[7] Institute of Physics, Faculty of Basic Sciences, École Polytechnique Fédérale de Lausanne, Lausanne, Switzerland

**Abstract**

Recently hBN has become an interesting platform for quantum optics due to the peculiar defect-related luminescence properties. Concomitantly, hBN was established as the ideal insulating support for realizing 2D materials device, where, on the contrary, defects can affect the device performance. In this work, we study the light emission properties of hBN flakes obtained by mechanical exfoliation with particular focus on extended defects generated in the process. In particular, we tackle different issues as the light emission in hBN flakes of different thicknesses in the range of hundreds of nm, revealing a higher concentration of deep level emission in thinner area of the flake. We recognize the effect of crystal deformation in some areas of the flake with an important blue-shift (130 meV) of the room temperature near band edge emission of hBN and the concurrent presence of a novel emission at 2.36 eV related to the formation of array of dislocations. We studied the light emission properties by means of cathodoluminescence and sub-bandgap excitation photoluminescence of thickness steps with different crystallographic orientations, revealing the presence of different concentration of radiative centers. CL mapping allows to detect buried thickness steps, invisible to the SEM and AFM morphological analysis.

**Introduction**

Hexagonal boron nitride (hBN) is a layered ultra-wide-band-gap semiconductor (5.95 eV indirect bandgap[1]) with a honeycomb lattice, where boron and nitrogen atoms alternate at the vertices of the planar hexagonal sp$^2$ structure. Recently hBN has become an interesting platform for quantum optics due to the peculiar luminescence properties of its point-defects.[2,3] Currently, different defects-related room-temperature single photon emitters (SPE) have been produced by various fabrication methods, like thermal annealing[4,5], ion implantation[6], strain field[7] and particle irradiation[5]. Using particles irradiation, electrons[5] and gallium ions[8] give rise mainly to the well-established zero-phonon-line at 630 nm, attributed to the carbon related complex[9,10]. Instead, neutrons[11] and ions (Nitrogen, Xenon and Argon)[12] irradiation generates the broad light emission at 850 nm due to the single boron-vacancy. Recently, a novel emission set at 435 nm has been generated by mean of electron beam irradiation with a micrometric localization across the hBN flake.[13] In parallel, hBN has gained enormous interest demonstrating to be the ideal insulating support for graphene[14–17] and transition metal dichalcogenides (TMDs) based electronic[18–20] and optoelectronic [21,22] devices, as well as an efficient encapsulating agent for environment-sensible 2D materials.[23–25] This important role is mainly due to its chemo-physical properties, such as the chemical internes, the atomic flatness and the absence of dangling bonds. In this regard, hBN flakes have become a mandatory component for the development of Van der Waals heterostructures[26] and 2D materials engineering techniques[27]. In particular, hBN has been widely used to decouple the 2D materials device from substrates, leading to large ameliorations of to the device performance compared to standard 300 nm thick $SiO_2$.[28] Albeit, recently it has been demonstrated that intrinsic defects in hBN can limit the field-effect performance of 2D materials device.[29] In addition, defects in hBN can introduce doping in gated devices when activated by various physical means, such as light exposure[30], electric field[31] and electron beam irradiation[32].

Therefore, the investigation of morphologically visible and invisible extended defects can considerably contribute to the understanding of the mechanisms affecting the performance of 2D materials devices on hBN substrates. The most comprehensive experimental tool to investigate and monitor the presence of optically active and non-radiative defects in hBN is cathodoluminescence (CL), i.e. the light emission stimulated by an electron beam excitation. CL experiments, standardly performed at cryogenic temperature, allows an easy excitation of the near band edge (NBE) emission of hBN[33], in the deep ultraviolet region, normally challenging to excite in photoluminescence (PL) experiments. In addition, CL hyperspectral mapping allows a nanoscale recognition of optically active and non-radiative extended defects. CL allows also a more straightforward detection of points- and extended defects related light emissions due to the high generation rate of phonons during the electron scattering events with the material lattice.[34,35]

In this work, we investigate the light emission properties of morphologically visible and invisible extended defects in the hBN crystal formed during the exfoliation process. By this means, we establish room temperature cathodoluminescence mapping as a convenient tool for a quick and reliable detection of extended defects in hBN. In particular, we explore the light emission properties of thickness steps with different crystallographic orientations, flake edges, and crystal deformation in hBN flakes obtained by mechanical exfoliation. We initially assess the presence of morphologically visible defects by atomic force microscopy (AFM) and scanning electron microscopy (SEM). Then we recognize the presence of morphologically invisible extended defects by means of panchromatic CL imaging. A careful comparative spectroscopic investigation by means of CL and sub-bandgap excited PL reveals a more intense blue emission from both the morphologically visible and buried thickness steps. In addition, we are able to recognize the effect of the crystal deformation on the hBN room temperature deep UV excitonic emission, a 130 meV blue-shift, and the concurrent appearance of an emission at 2.36 eV, mainly localized as a line feature at the edge of the deformed areas of the crystal.

**Results Discussion**

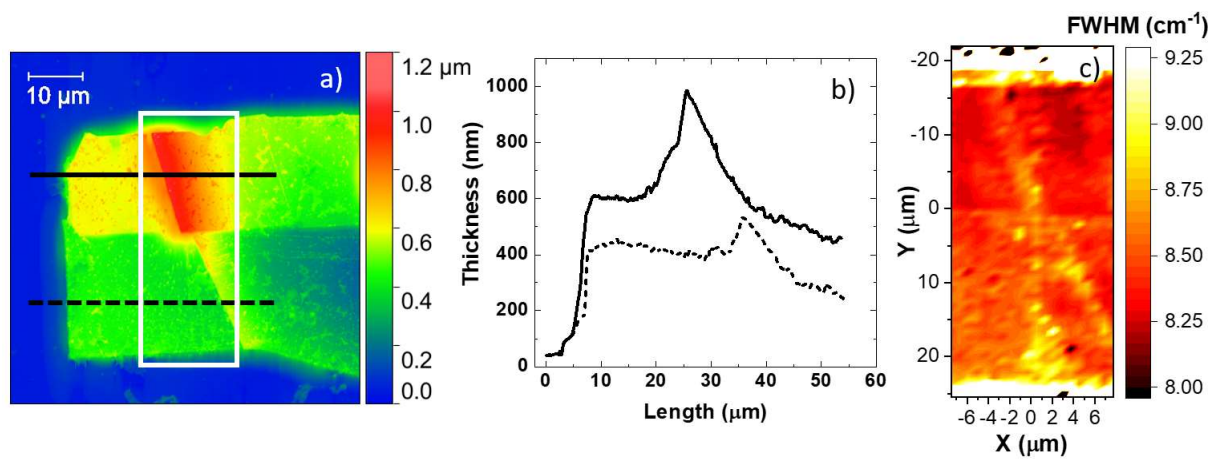

**Fig.1 a) AFM morphological analysis of the hBN flake b) line profiles obtained along the black full and dashed lines. c) Raman map of the FWHM of the hBN $E_{2G}$ Raman mode. The Raman map is acquired in the region highlighted by the white square in panel a).**

Figure 1a presents the AFM analysis of the hBN flake, obtained by mechanical exfoliation (see Experimental Section). The morphological analysis reveals that the flake is divided in four quadrants with different thickness. The upper-left quadrant has an average thickness of 605 nm, while the thickness of the upper right quadrant is 500 nm. The lower quadrants have a lower thickness of 430 nm (left) and 250 nm (right). In the horizontal transition among the different quadrants a strong upward deformation of the hBN crystal is evident for both the upper and lower regions of the flake. The AFM line profile (Fig. 1b), reveals that the local thickness reach the maximum value of 985 nm and of 534 nm in case of the upper and lower regions, respectively. The deformation of the crystal is an upward bending of the flake, with a protruding region width of 20 μm and 17 μm for the upper and lower parts, respectively. In addition, the AFM analysis highlights the presence of different thickness step in the hBN flake; in particular, the thickness step in the upper part of the flake has variation of thickness of 187 nm and an angle of 15° with respect to the horizontal line of division among the upper and lower part of the flake. While, in the lower part of the flake, the AFM profile reveals the presence of two thickness step with different angle and thickness variation, one perpendicular to the horizontal line of breaking with a thickness of 38 nm and the second with an angle of 23° and thickness variation of 126 nm.

In order to assess possible local variations of the crystallinity due to the presence of the morphologically visible extended defects, such as thickness step and deformed areas, Raman spectroscopic mapping is carried out. The Raman peak which is conventionally investigated in hBN is at 1366 cm$^{-1}$. This Raman mode is attributed to the in-plane atom vibrations ($E_{2g}$ mode).[36] The benchmark for the crystallinity of hBN is the full width half maximum (FWHM) of such peak. The standard FWHM of the hBN, employed in this study is $\approx$ 9 cm$^{-1}$, as previously reported.[33] The Raman map of the FWHM of the hBN Raman mode, reported in Figure 1b, evidences that the average FWHM is different for the different quadrants of the flake, depending on the thickness of the hBN. The two upper quadrants with a thickness ranging from 500 nm to 600 nm have an average FWHM of 8.3 cm$^{-1}$, meanwhile the lower quadrants with a thickness varying between 300 nm to 500 nm have an average FWHM of 8.5 cm$^{-1}$. A previous work, comparing the FWHM of hBN from different sources, reported that hBN, similar to the one employed in this work, presents a FWHM of 7.3 cm$^{-1}$ with a thickness of tens of nanometers.[37] Therefore the thickness appears to have an impact on the Raman mode FWHM of hBN. In the case of the morphologically visible extended defects the FWHM of the Raman modes increases up to 9.3 cm$^{-1}$. The thickness step of the upper quadrant has a FWHM of 8.6 cm$^{-1}$ in line with the average value of the area. A similar value is observed for the deformed crystal, demonstrating that the deformation does not affect the crystallinity of hBN. The thickness step of the lower quadrants presents a FWHM of 8.9 cm$^{-1}$ and 9.3 cm$^{-1}$. It is worth noting that the average value for the FWHM measured at the edge of the flake is 9 cm$^{-1}$. Therefore, we can consider the flake edge as an extended defects similar to the thickness steps. The Raman spectra of the areas, highlighted by the map, and the histogram of the hBN Raman mode FWHM are reported in Figure S1 of the Supporting Information.

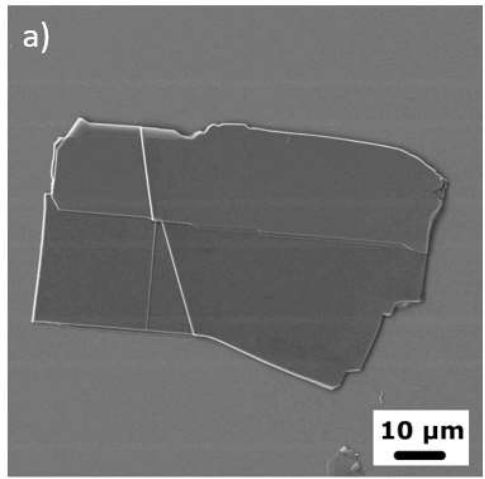

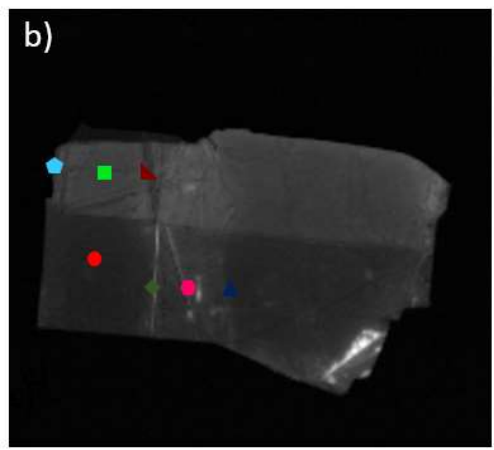

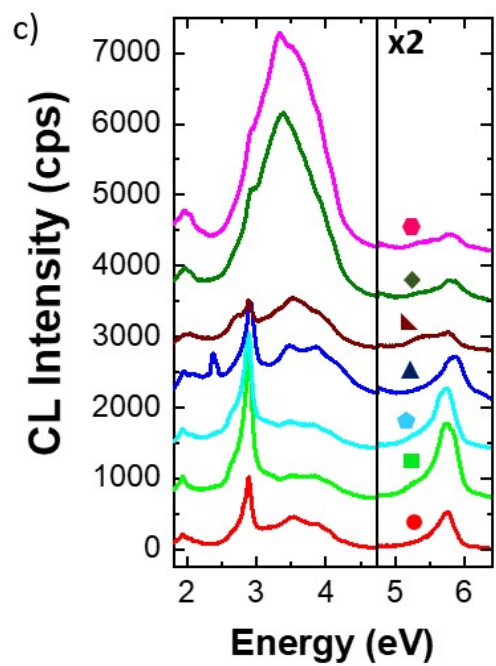

**Figure. 2 a) SEM micrograph of the flake in analysis, b) the panchromatic map of the same flake where the colored dots represent the point where the spectra of panel c) is**

**acquired. c) CL spectra of the pristine hBN areas with different thickness and of extended defects. The color code of the spectra is the same of the dots of panel b). ● pristine hBN (378 nm thick), ■ pristine hBN (575 nm thick), ⬠ flake edge, ▲ deformed crystal area, ◤ dark slanted thickness step, ◆ bright vertical thickness step, ● bright slanted thickness step. The NBE range is multiplied by 2 for clarity.**

The SEM imaging (Fig 2a), using a secondary electron detector, reveals mainly the same features highlighted by AFM. The SEM analysis shows the presence of the different thickness steps, albeit it allows a more straightforward recognition than AFM of the vertical thickness step in the lower quadrant with a thickness of tens of nanometers. The panchromatic CL map (PanCL) is presented in Figure 2b. The PanCL signal corresponds to the emitted photons integrated across the whole detection range of the system, in this particular case from 6 eV (200 nm) down to 1 eV (1100 nm). The PanCL map allows detecting the presence of morphologically invisible extended defects. In fact, in the PanCL map it is possible to observe an intense emission from the lower right corner of the flake, in an area that does not present any features in the SEM morphological micrograph. Moreover, it is possible to detect some dark contrast line, one is near the upper left corner of the flake and additional ones are at the base of the deformed crystal areas. These dark contrast lines are probably due to the mechanical stress applied on the flake during the exfoliation process.[38] In addition, the PanCL map allows to understand the recombination nature of the extended defects. In fact, defects with radiative recombination properties will have a more intense light emission and thus a brighter contrast while non-radiative recombination defects will have a dark contrast with respect to the pristine areas of the flake. For instance, the thickness step of the lower quadrant presents a bright contrast in the PanCL map, while the thickness step of the upper quadrant presents a dark contrast with respect to the pristine parts of the flakes. Using the standard formula for an object PanCL contrast ($C$), we can evaluate a more quantitative analysis of the defects light emission nature[39,40]:

$$C = \left(\frac{I_{OBJ} - I_{BG}}{I_{BG}}\right) \times 100$$

where $I_{OBJ}$ is the PanCL intensity of the of the feature of interest, while $I_{BG}$ is the average Panchromatic CL emission of background, in this particular case the PanCL emission of pristine hBN (red and green dots, considering the different thickness of hBN in Fig. 2b). In this case dark contrast objects have a negative *C* while features with a higher light emission present a positive *C*. In the case of the thickness steps, in the upper quadrant the thickness step has a PanCL contrast of -14%, while the vertical and slanted thickness steps in the lower part of the flake have a contrast of 117% and 38% respectively. This clearly demonstrates that similar defects, as the thickness steps, can have totally different radiative recombination properties. This can be due to different crystallographic orientation of the breaking line of the crystal.[41] An intense emissive area is present in the lower right corner of the flake, with a PanCL contrast of 560%, without presenting any morphological feature in the SEM image. In order to have additional insights, CL spectroscopy has been carried out in the sites indicated by colored dots in the PanCL map of Figure 2b. The same color code has been employed for the CL spectra of Figures 2c. The CL spectra of the pristine areas of the hBN flake with different thickness are shown in Fig. 2c. The pristine hBN spectrum is composed of different sharp peaks and broad bands. The sharp peak at 5.76 is related to the room temperature near band edge phonon assisted recombination of the hBN.[42] This emission at room temperature is the convolution of *s*-like free exciton related emissions.[37,43,44] The tail on the low energy side at about 5.38 eV is attributed to the D series, radiative recombination due to the presence of extended structural defects as stacking faults[45], dislocation[46] or grain boundaries[44]. The spectrum, then presents a broad band composed of different peaks between 4 eV and 3 eV, which identification is still unclear. In fact, this luminescence was firstly erroneously attributed to a band-gap at 4 eV[47,48] while it was then attributed to radiative transition due to impurities[43]. This band is often defined as deep defects luminescence. Nowadays, it is still under debate if this light emission is due to impurities or structural defects. In addition, a recent work reported the twist-angle dependent enhancement of the UV peaked emission at about 4.1 eV (300 nm) in hBN twisted thick homo-structures.[49]

The sharp peak at 2.88 eV (430 nm) is due to the second order of the NBE emission. This peak should not be confused with the sharp emission at 2.84 eV (436 nm) due to electron irradiation induced defects.[13,38] Finally, the spectrum presents an emission at 1.92 eV (646 nm) most likely due to the second order of the deep defect emission. Considering the CL spectra reported in Figure 2c, we compare the CL spectra of 378 nm (red line) thick and 535 nm (green line) thick pristine areas of the hBN flake, demonstrating that the hBN thickness has an important influence on CL emission. First the NBE peak is less intense in the thinner area of the flake. The less intense emission of the 378 nm thick area can be erroneously attributed to the less amount of material of such area. In this regard we have to keep in mind that the CL analysis is carried out at 5 kV. Therefore, the maximum penetration depth of the electrons is about 350 nm as shown in Figure S2 of the Supporting Information. Figure S2 presents also the generation of the CL signal as function of the depth, revealing that the generation volume is completely inside the hBN flake for both the 378 nm and 535 nm thicknesses.[50] Therefore, we cannot assign the lower intensity of the NBE emission in this area to the lower thickness probed during the CL experiments. However, we have to take into account that the broad band due to the deep defects has a similar intensity for both the areas, with a more intense peak at 3.48 eV for the thinner area. Therefore, lower intensity of the NBE emission can be due to a higher concentration of defects, giving rise to the 3.8 eV band, in the thinner part of the flake, thus causing a decrease of the NBE radiative recombination. On the contrary, the 535 nm thick area of the hBN flake presents a more pronounced shoulder on the low energy side of the NBE emission, that is not detected in the 378 nm thick area of the flake.

We carried out CL spectroscopy also on both the morphologically visible and invisible extended defects. The spectrum of the flake edge (cyan line) does not present any significant difference from the 535 nm thick pristine area. Instead the spectrum of the crystal bend (blue line in Fig. 2c) presents some interesting variations: i) the NBE peak presents an important blue shift, being peaked at 5.87 eV, with a concurrent broadening of such emission; ii) the deep defects band presents an enhanced intensity of all its components; iii) in this spectrum there is the appearance of a sharp band at about 2.36 eV, which is not present in any other spectrum acquired on the flake. This emission has been previously reported, owning a single photon character, and it has been attributed to strain effect affecting the crystal lattice.[7] This previous attribution is supported by the blue-shift of the NBE emission, reported in this work, because such effect is caused by strain. Similar localized emission was previously reported in Ref. [38], with a large wavelength range integration (482-735 nm) of the CL emission and it was related to dislocations which are assembled into arrays and aligned along definite crystalline directions.[46]

Considering the different thickness steps of the flake, the one of the upper quadrant with a dark contrast in PanCL map presents a spectrum (dark red line) where the NBE emission is deeply quenched and the shape of the deep defects band is different with respect to the pristine hBN. In fact, the band is peaked at 3.5 eV with a clear suppression of the other high energy components (3.88 eV and 4.1 eV). This indicates that the thickness step has a majority of structural defects giving rise to this band at 3.5 eV with respect to the defects emitting at 3.88 eV and 4.1 eV. This hypothesis is supported by CL spectra of the thickness steps of the lower quadrant (with a bright contrast in the PanCL map), where the intensity of this band peaked at about 3.5 eV is six times more intense with respect to the pristine hBN areas. In fact, the thickness step spectra in the lower quadrant (dark green and magenta lines) present a similar quenching of the NBE emission probably due to the competitive radiative process of the deep defects band that is more strongly intense in such defects.

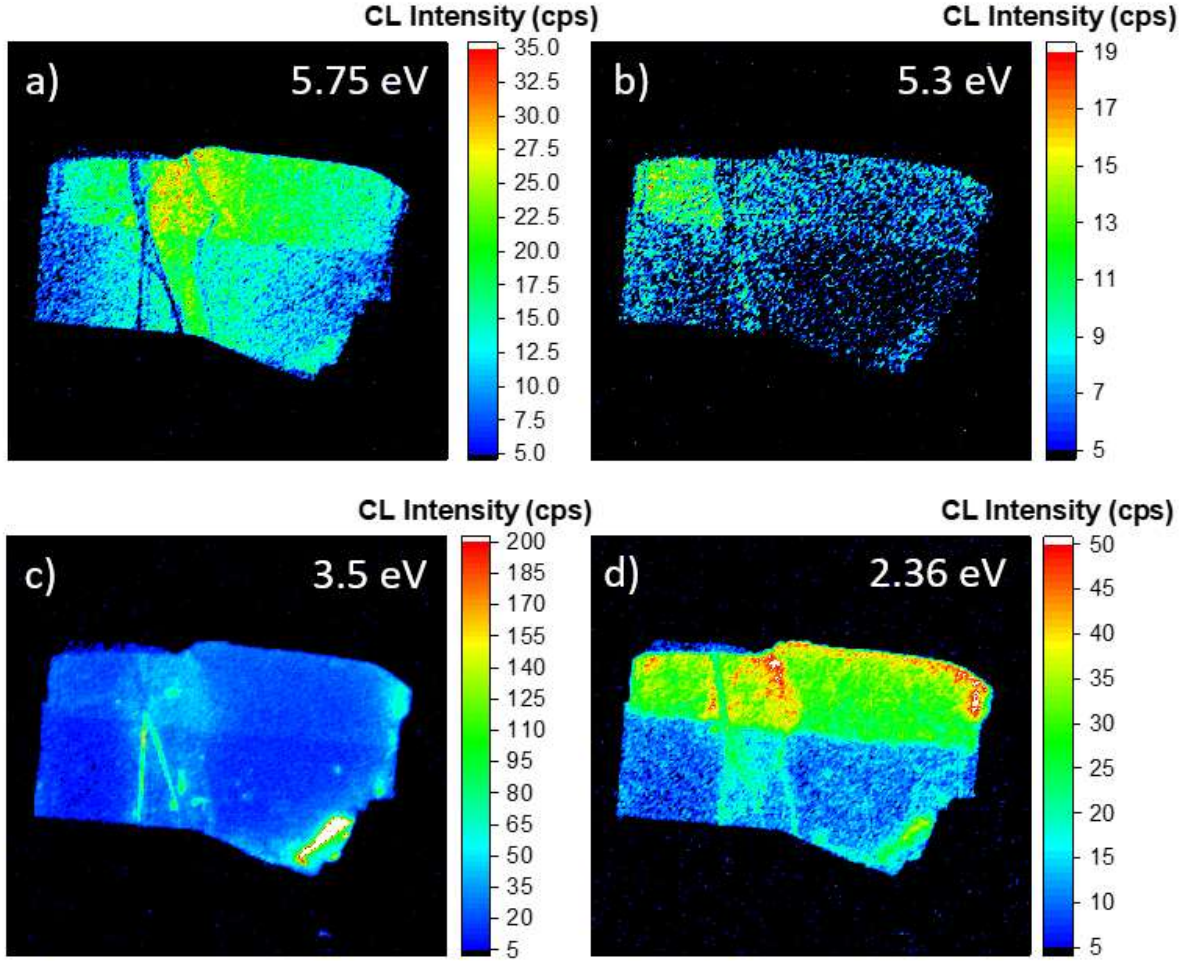

**Figure 3 CL mapping of the principal emission in hBN flake: a) 5.75 eV attributed to NBE, b) 5.3 eV related to the low-energy tail of NBE emission, c) and d) deep defects emission at 3.5 eV and the emission at 2.36 eV attributed to crystal deformation.**

In order to obtain a clear view of the CL emission localization, we carried out CL hyperspectral mapping, reported in Figure 3. Figure 3a presents the monochromatic CL (MonoCL) map of the NBE emission at 5.75 eV. This emission is more intense in the upper part of the flake where the flake is thicker. The maximum intensity is obtained in the areas where the hBN flake deforms. In addition, it is worth noting that all the thickness steps present a dark contrast, resulting in a quenched NBE emission at such defects. Moreover, as highlighted by the CL spectra of Fig. 2b, additional dark contrast lines appear next to the crystal deformation areas due to the blue-shift of the NBE emission in such areas. In order to evaluate the localization of the D peak, the map of the low energy tail of the NBE emission, at 5.3 eV, is reported in Fig.

3b. This emission is localized mainly in the thicker part of the flake, indicating a clear role of the flake thickness. In order to evaluate possible difference in the deep defects band, Figure 3c shows the MonoCL map obtained at 3.5 eV, respectively. The bottom-right corner presents an intense emission at such energies revealing a highly defective area. The MonoCL map of the emission at 2.36 eV is presented in Fig 3d. This emission is localized mainly in deformed crystal areas. This map highlights the formation of linear features at the starting point of the deformation areas. These linear features were previously assigned to array of dislocation formed during the mechanical exfoliation of hBN crystals.[44] Figure S3 present a high magnification CL mapping of the defective area in the bottom right region.

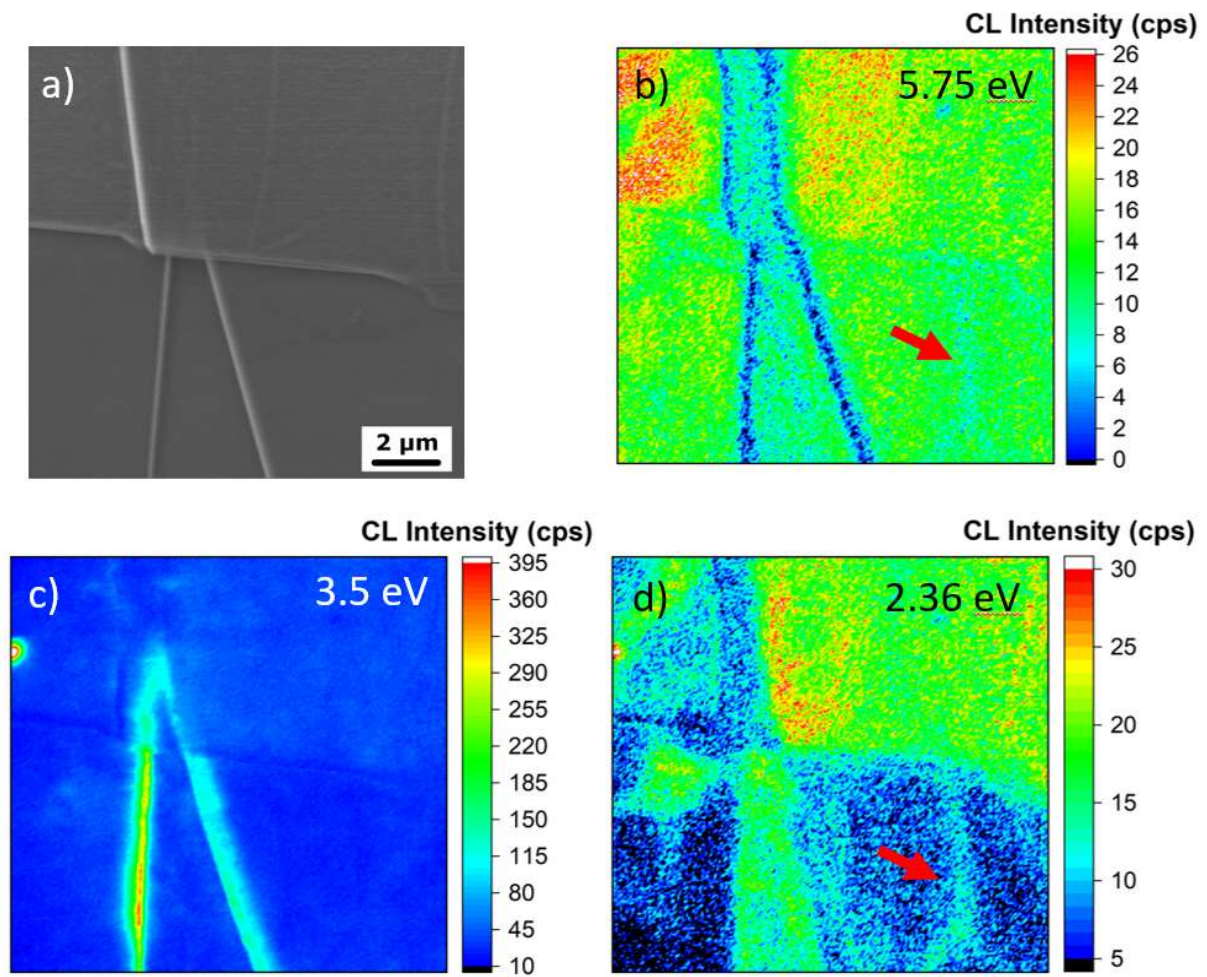

**Figure 4 High magnification CL mapping of the hBN flake center. a) SEM micrograph. b) 5.75 eV CL map. c) 3.48 eV CL map. d) 2.36 eV CL map**

The high magnification CL mapping of the center of the flake reveals interesting additional insights. The SEM micrograph (Fig. 4a) shows more details of the morphology of the different thickness steps. From the monoCL map at 5.75 eV (Fig. 4b), all the thickness steps present a quenching of the NBE emission of hBN. However, it is worth noting that this CL map allows to track the thickness step that extends underneath the upper part of the flake. This result is also confirmed by the monoCL maps obtained at 3.5 eV where the presence of the buried thickness step is highlighted by the intensity increase of this emission. In addition, this monoCL map also shows a quenching of the intensity where the AFM has evidences the crystal deformation, as highlighted by a red arrow, the same feature presents a more intense emission in the 2.36 eV monoCL map, as shown in Fig. 4d. These two findings demonstrate that CL mapping is able to detect even morphologically invisible extended defects in hBN. Further analysis of 50 nm and 500 nm thick flakes are added in Figure S4 and S5 respectively.

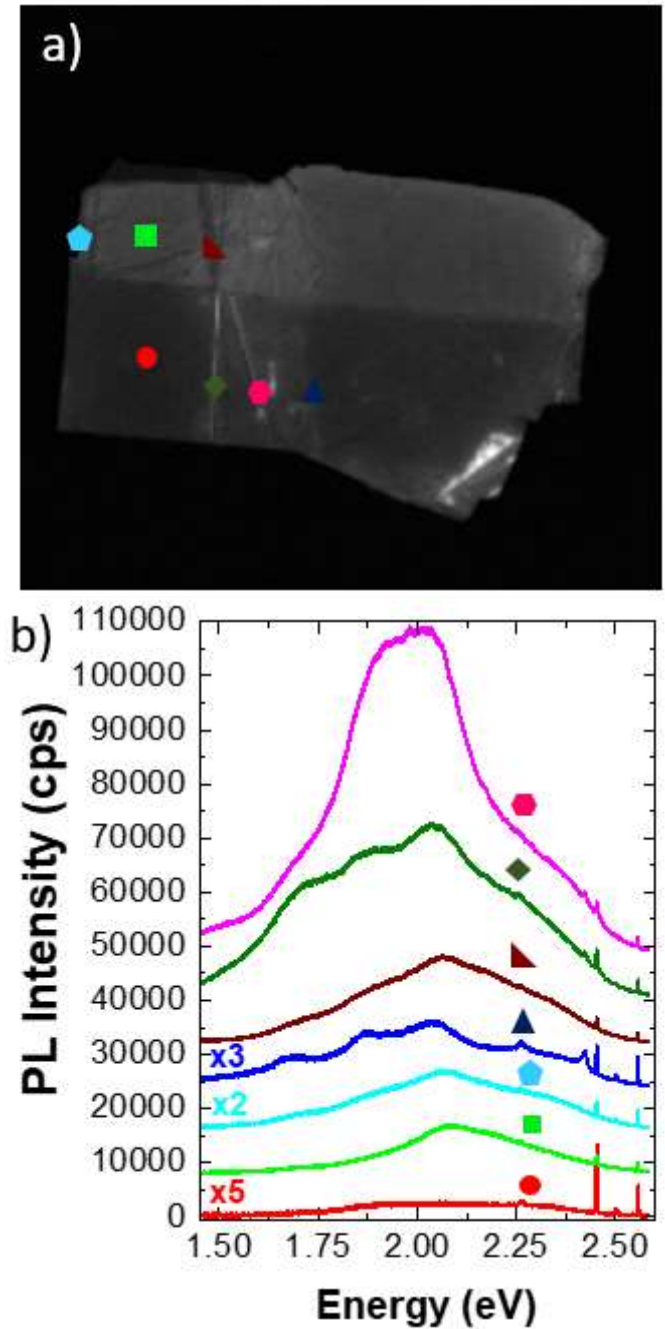

**Figure 5.** Sub-bandgap excitation photoluminescence analysis: a) PL spectra of ● pristine hBN (378 nm thick), ■ pristine hBN (575 nm thick), ⬠ flake edge, ▲ deformed crystal area, ◤ dark slanted thickness step, ◆ bright vertical thickness step, ⬢ bright slanted thickness step. Some curves have been multiplied by the factor indicated in the figure.

.

To obtain additional insights on the optical properties of the morphologically visible and invisible extended defects we carry out photoluminescence spectroscopy with a blue laser excitation (2.63 eV, 473 nm), therefore a sub-bandgap excitation. It is worth noting that the color code of the spectra is the same for CL and PL spectroscopic analysis as well as the dots highlighted in the PanCL map, that we have reported as Fig. 5a. Figure 5b show the PL spectra of the pristine area of the hBN with a thickness of 378 nm (red line) and 535 nm (green line). As suggested by the CL spectroscopic analysis of Fig. 2c, the pristine area with lower thickness presents a broad emission set at 2.09 eV with a shoulder on the low energy side at about 1.7 eV, with similar behavior of the CL data regarding the deep defects band at about 3.8 eV. The pristine area with a 535 nm thickness presents faint broad emissions at 2.36 eV and 1.84 eV with an intensity twenty times lower than the area with a lower thickness. The sharp set of lines at 2.56 eV and 2.45 eV is attributed to vibrational peak of the silicon substrate and of the hBN, respectively. It is worth mentioning that, in comparison with the CL analysis, the PL spectrum acquired on the 535 nm thick area edge (Fig. 5b, cyan line) presents a broad band peak at 2.06 eV with an intensity twice as high with respect to the pristine area. In addition, it is possible to define the components of such broad band. In fact, there is one clear peak on the high energy side at about 2.3 eV and two other peaks on the low energy side set at 1.9 eV and 1.7 eV. It is worth highlighting the peak appearing at 2.26 eV. The other spectrum presented in Fig. 5b is acquired in the deformed crystal area (blue line). The lineshape of the PL spectrum is similar to the spectrum acquired on the flake edge; albeit the main peak of is slightly red-shifted, being peaked at 2.03 eV, and the additional peaks at 1.87 eV and 1.62 eV are more pronounced. Moreover, two additional sharp PL peaks at 2.42 eV and 2.26 eV appears in the area of the crystal deformation. This is in good agreement with the spectroscopic CL results reported in Fig. 2. Figure 5b also reports the PL spectra of the different thickness steps recognized on the flake. In case of the thickness step with a dark contrast in the PanCL map (dark red line), the PL spectrum presents a broad band centered at 2.06 eV, similar to the spectral of the 378 nm thick pristine area but with a slightly higher intensity. In case of the thickness steps with a bright contrast in the PanCL map, the PL band is still peaked at 2.06 eV, but presents an intensity that is three times and six times higher for the vertical and slanted thickness steps respectively (dark green and magenta lines). The lineshape of these spectra is slightly different. For the vertical thickness step of the lower part of the flake, additional PL peaks form two clear shoulders at 1.87 eV and 1.72 eV. The intensity of the 1.87 eV becomes even more prominent for the slanted thickness step of the lower part of the flake, where the main peak broads drastically due the convolution of the 2.06 eV and 1.87 eV peaks.

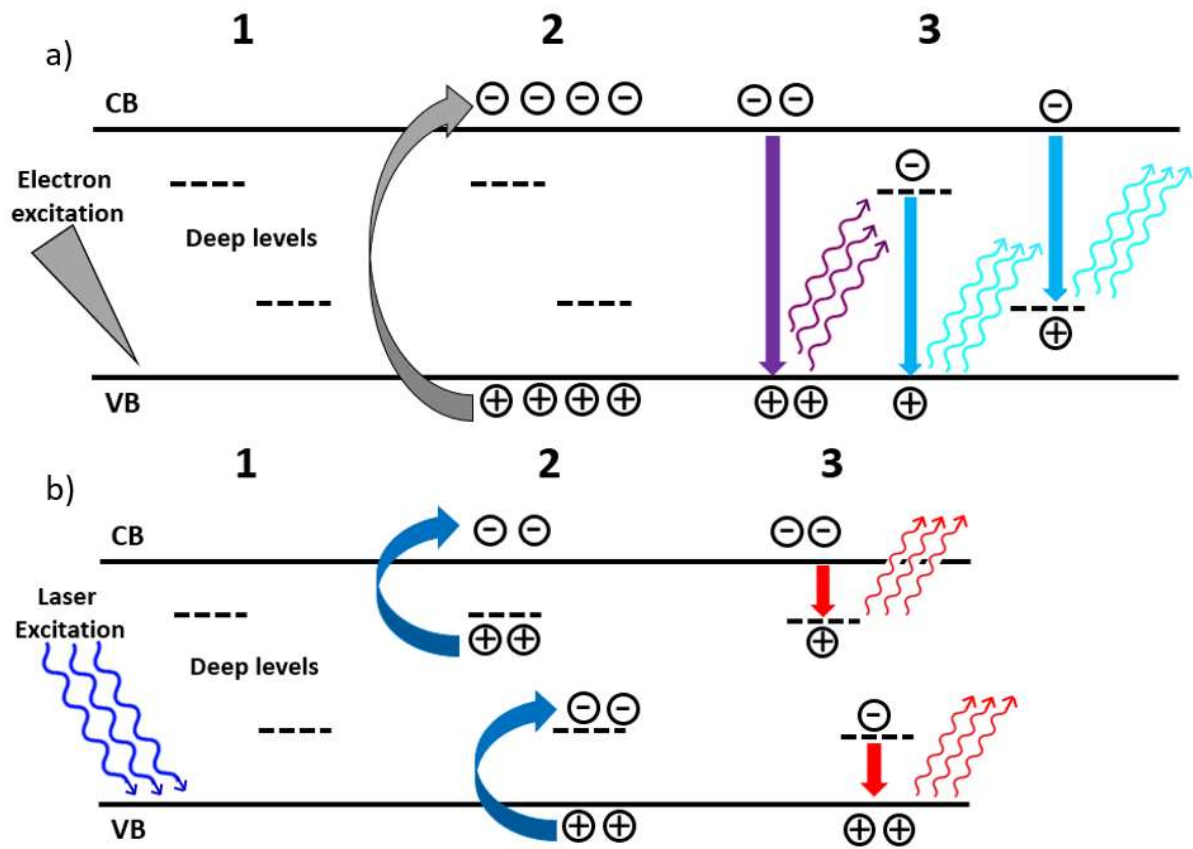

**Figure 6 Sketch of the excitation regimes in CL (a) and PL (b) experiments.**

The similar behaviors of the CL and PL emissions can be explained attributing the emission in the two different experiments to the same radiative center but with two different excitation regimes. In order to clarify the main differences, simplified sketches of the CL and PL excitation regimes are presented in Fig. 6. Figure 6a presents the CL excitation, where the electrons excited by the incident electron beam, are promoted from the valence band to the conduction band of hBN. This excitation regime allows to have band-to-band radiative recombination and, in case of the presence of deep levels in the bandgap, additional emissions related to the radiative recombination at these intra-bandgap states. In case of the sub-bandgap excitation regime of the PL experiments (Fig. 6b), the band-to-band excitation is not allowed and therefore the possible excitation paths are: the excitation from the valence band to a deep level (with an energy distance from the valence band similar to the one of the excitation photons) and the excitation from a deep level to the conduction band. In this regard, hBN emission can have energy lower than the excitation photons, and therefore, in the case of the hBN lower than the mid bandgap energy. In this picture we can consider a possible approximated evaluation to compare the behaviors of the extended defects related emissions in the CL and PL experiments. The evaluation can be resumed as follows:

$$E_{G(hBN)} \approx h\nu_{CL} + h\nu_{PL}$$

The energy of the CL emission $h\nu_{CL}$ is summed to the energy of the PL emission $h\nu_{PL}$ and it should result in an energy value ($E_{G(hBN)}$) close to the hBN bandgap energy.[1] This evaluation is strongly approximated because it does not take in account the actual material band structure (i.e. recombination at different $k$ space points) and the shape of the bands of deep levels, considering directly in a dispersion-less band regime.

Moreover, this evaluation neglects the phonon energy necessary for an indirect radiative recombination in case of cathodoluminescence emissions. However, under these assumptions, we can attribute the 3.8 eV CL emission to the 2.1 eV PL emission to the same intra-bandgap state, as suggested by the data of the pristine hBN with different thickness. Table 1 resumes all the correlation among CL and PL analyses, and compares the previous attribution of the emission with the new insights gained in this work.

**Table 1**

| CL Emission Energy (eV) | PL Emission Energy (eV) | **Previous attributions** | **Insights gained in this work** |
|---|---|---|---|
| 5.75 | Not present | Room-temperature NBE of hBN, S bands [33,43] | Intensity dependent on the thickness and blue-shifted in |

|       |     |                                                                                                      | deformed regions                                                                                                   |
| ----- | --- | ---------------------------------------------------------------------------------------------------- | ------------------------------------------------------------------------------------------------------------------ |
| 4.1   | 1.8 | Impurities or structural defects[43,51]                                                              | Mainly PL intensity increase in presence of a thickness step, most likely due to structural defects.               |
| 3.88  | 2.1 | Phonon replica of 4 eV emission [51], donor acceptor pair with strongly localized acceptor [52].     | CL and PL intensity increase in presence of a thickness step, most likely due to structural defects.               |
| 3.48  | 2.4 | Phonon replica of 4 eV emission [51]                                                                 | No particular evidences                                                                                            |
| 2.36  | 2.3 | Array of dislocations [46], strain related effects with SPE character[7]                             | Emission appear at the edge of deformed areas and related to strain effect.                                        |

The main findings of this study are resumed in the following. The NBE emission at 5.75 eV is blue-shifted by 130 meV with a concurrent broadening in the deformed region of the flake. Therefore, this shift of the emission can be related to strain effect, albeit previous works on deformed hBN presented mainly a red shift of the UV emission[45]. The complex band peaked at 3.88 eV shows an intensity enhancement in the case of the analysis of the thickness steps, and therefore it is possible to attribute this deep level emission to structural defects formed during the exfoliation process. This behavior is confirmed by the PL experiments. This emission is a complex feature with different components at 4.1 eV and 3.48 eV. These bands where previously attributed to the phonon replicas of the 4 eV emissions[51], albeit both the CL and PL experiments seems to assign these emissions to different origins, most likely structural defects, because the enhancement of the different components is not homogeneous with the evident change of the CL and PL spectra lineshape in the analysis of the different thickness steps, see Fig. 2c and Fig. 3c. Previously a broad and intense luminescence band set at 3.9 eV was assigned to donor-acceptor pair, involving a strongly localized acceptor complex center.[52]

In addition, we can attribute a novel emission, detected both in CL and PL experiments, at 2.3 eV to strain induced structural defects, previously attributed to array of dislocations[38,46], suggested by the localization of such emission mainly in crystal deformation regions.

**Conclusions**

In this work, we tackle different issues regarding extended defects in hBN formed in the mechanical exfoliation process by means of CL and sub-bandgap excitation PL experiments. We report the light emission of hBN flake with different thickness in the range of hundreds of nm, revealing a higher concentration of deep level, giving rise to the visible emission at 3.48 eV, in thinner area of the flake. We study the light emission properties of thickness steps with different crystallographic orientations, which allows different light emitting properties with both electron and photons excitations, revealing a possible attribution of 3.44 eV band (2.1 eV in PL) to structural defects. Moreover, CL mapping allows to detect buried thickness steps, invisible to the SEM and AFM morphological analysis. Finally, we recognize the effect of crystal deformation in particular areas of the flake with an important blue-shift (130 meV) of the room temperature NBE emission of hBN and the concurrent presence of a novel emission at 2.36 eV in both CL and PL experiments, related to the formation of array of dislocations.

**Experimental**

Nowadays the main technique to obtain high quality hBN flakes with lateral size of hundreds of micron is still the mechanical exfoliation of single crystals obtained by high pressure technique. [43,53] We obtain the hBN flakes via standard micro-mechanical exfoliation[54] from bulk single crystals[43]. Peeling with adhesive tape (Nitto Blue Tape) is limited to 10-20 repetitions. The $SiO_2$/Si exfoliation substrate is cleaned by oxygen plasma prior to deposition of the tape and heated to 50°C during few minutes to promote adhesion of the flakes. After cooling to room temperature by air convection, the tape is slowly peeled off with the help of a micromanipulator, and the samples are cleaned overnight in organic solvents to remove possible glue residuals.

AFM measurements are performed using a Dimension Icon AFM (Bruker) operating in Peak Force mode and using ScanAsyst air probe.

Raman and photoluminescence (PL) spectroscopies are carried out with a Renishaw InVia system, equipped with confocal microscope, a 473 nm excitation laser and a 2400 line/mm grating (spectral resolution 0.5 nm). The Raman analyses are performed with the following parameters: excitation laser power 500 µW, acquisition time for each spectrum 1 s, spot size 800 nm thanks to a 100X objective (NA= 0.85). The PL spectroscopic experiments are carried out with the same parameters except for the acquisition time that is raised up to 10 s.

The cathodoluminescence experiments are carried out in an CL dedicated Attolight Rosa SEM microscope.[55] The e-beam current is about 91 pA with an accelerating voltage of 5 kV. The CL signal is sent to a spectrometer with 32 cm of focal length by means of an objective (NA= 0.71) placed in the electron microscope. The system is equipped with a Peltier-cooled charge-coupled device (CCD) and 600 l/mm diffraction grating. The SEM images are obtained with a standard Everhart–Thornley detector for secondary electrons.

**Acknowledgements**
C.G., N.T., A. F. I M. and F.F. want to thank the Interdisciplinary Centre for Electron Microscopy (CIME-EPFL) for the support. S.P. and F.F. want to thank Dr. Camilla Coletti and her group for the scientific discussions. C.B., N.T. and A. F. i M. acknowledge the funding from SNSF for projects nr 40B2-0_176680 and 200021_196948. K.W. and T.T. acknowledge support from the Elemental Strategy Initiative conducted by the MEXT, Japan (Grant Number JPMXP0112101001) and JSPS KAKENHI (Grant Numbers 19H05790, 20H00354 and 21H05233).

Supporting Information

# Luminescence of mechanical exfoliation induced extended defects in hexagonal boron nitride with variable thickness.

# Luminescence of morphologically visible and invisible extended defects in mechanically exfoliated hBN flakes

G. Ciampalini[1,2,3]*, C. V. Blaga[4]*. N. Tappy[4], S. Pezzini[3], Watanabe[5], Taniguchi[6], F Bianco[3], S. Roddaro[1,3], A. Fontcuberta i Morral[4,7], F. Fabbri[3]

*These authors contributed equally to the manuscript.

[1] Dipartimento di Fisica E. Fermi, Università di Pisa, Largo B. Pontecorvo 3, I-56127 Pisa, Italy

[2] Graphene Labs, Istituto Italiano di Tecnologia, Via Morego 30, I-16163 Genova, Italy

[3] NEST, Istituto Nanoscienze – CNR, Scuola Normale Superiore, Piazza San Silvestro 12, 56127 Pisa, Italy

[4] Laboratoire des Matériaux Semiconducteurs, Institute of Materials, Faculty of Engineering, École Polytechnique Fédérale de Lausanne, 1015 Lausanne, Switzerland

[5] Research Center for Functional Materials, National Institute for Materials Science, 1-1 Namiki, Tsukuba, 305-0044, Japan

[6] International Center for Materials Nanoarchitectonics, National Institute for Materials Science, 1-1 Namiki, Tsukuba 305-0044, Japan

[7] Institute of Physics, Faculty of Basic Sciences, École Polytechnique Fédérale de Lausanne, Lausanne, Switzerland

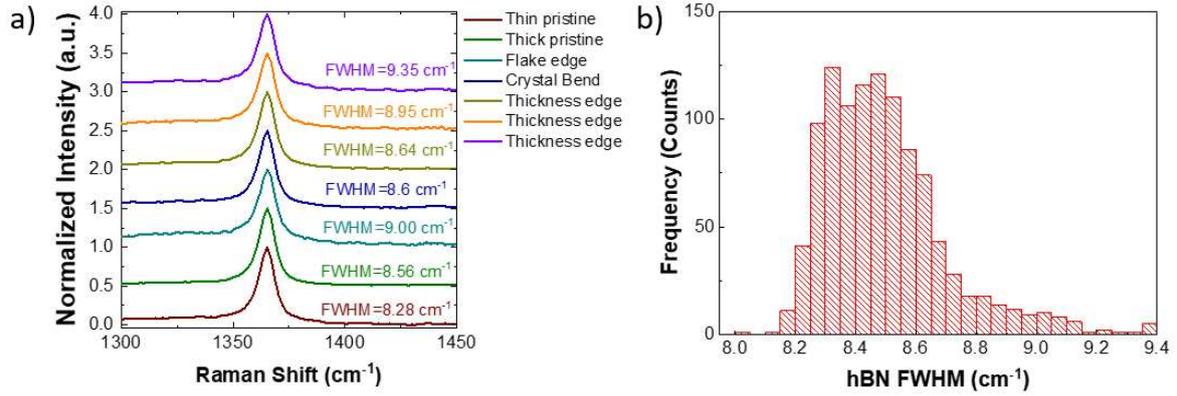

**Figure S1 a) Raman spectroscopy of the same location analyzed by CL spectroscopy in Figure 1 d). b) Distribution of E2g mode FWHM obtained from the Raman map of Figure 1b).**

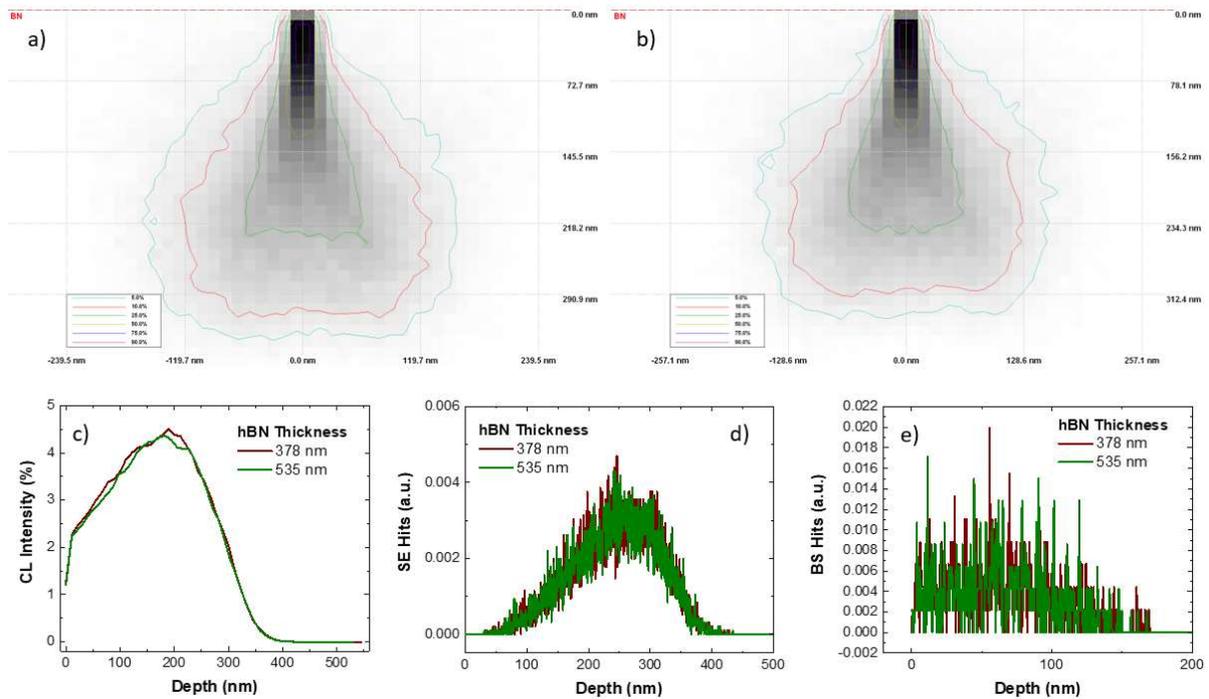

**Figure S2 Montecarlo simulation of the electron beam interaction with hBN. a) Energy release map in a 378 nm thick hBN. b) Energy release map in a 535 nm thick hBN. c) Depth profile of the CL emission of the hBN. d) Normalized scattering events depth profile. g) Depth profile generation of backscattered electrons.**

This comparison evidences that the electron energy release is similar for the different hBN thickness.

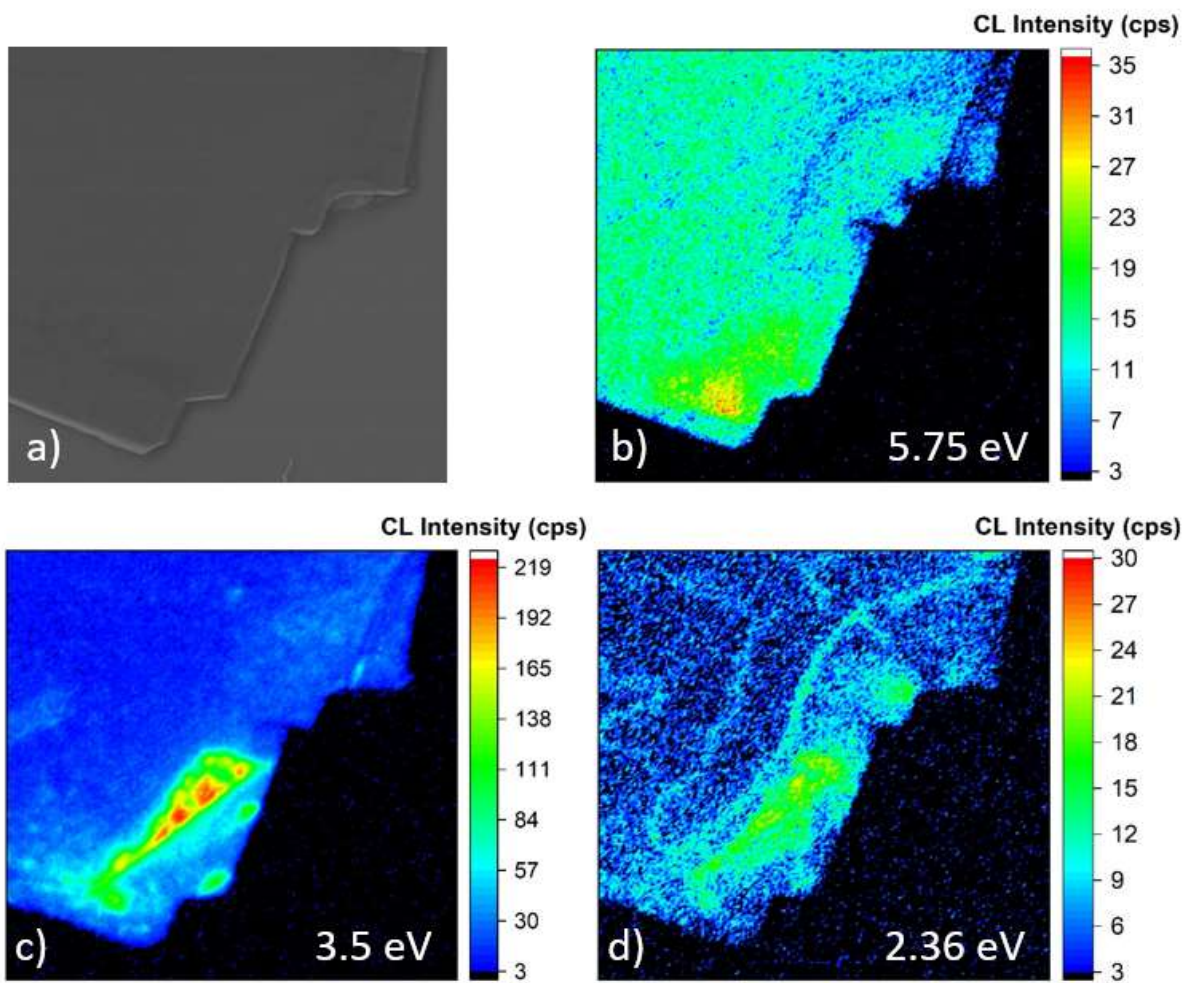

**Figure S3 High magnification CL mapping of the defective area at the bottom right corner of the analyzed hBN flake.**

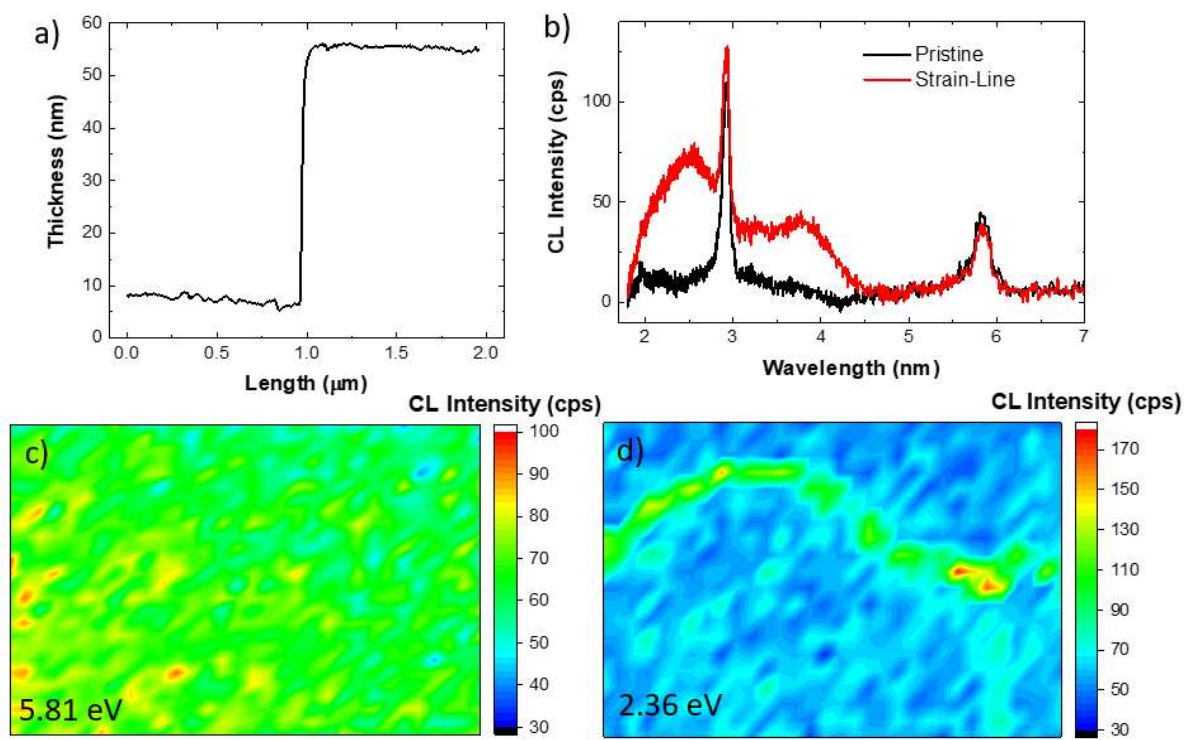

**Figure S5 CL analysis of a 50 nm thick hBN flake. a) Thickness line profile obtained by AFM. b) CL spectra obtained in a pristine area and in an area with a clear contrast in CL maps. The comparison results in an more intense emission at 3.8 eV and 2.6 eV with a similar 5.75 eV emission intensity. c) MonoCL map of the 5.75 eV emission related to the hBN NBE. d) MonoCL map of the 2.36 eV emission, highlighting the line feature where strain induced emission is more intense.**

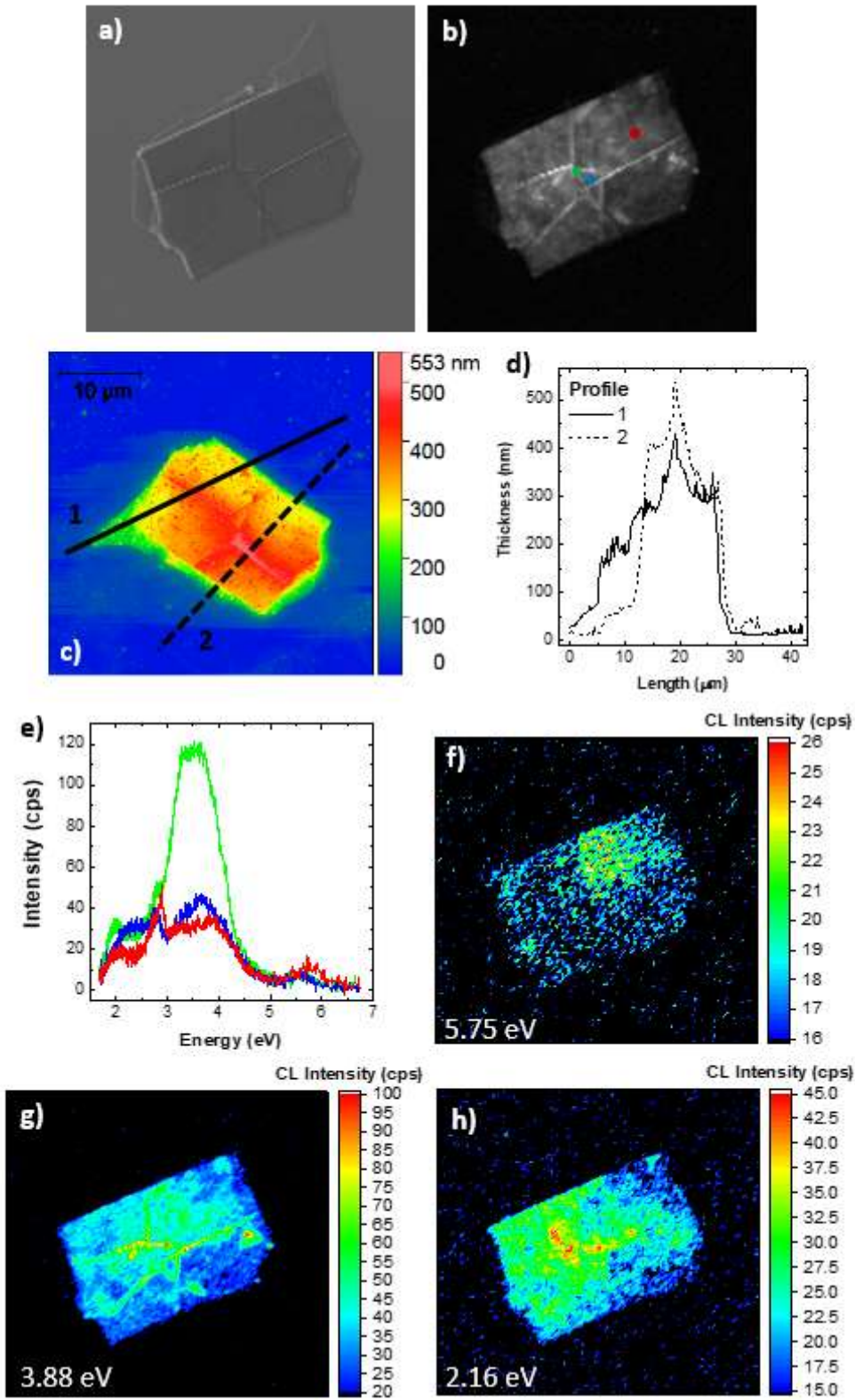

**Figure S6 CL analysis of a 500 nm thick hBN flake. a) SEM micrograph. B) Pan-CL maps where the colored spots indicate where the spectra of panel e are acquired. c) AFM thickness map. The full and dashed lines represent the area where the line profiles are**

**acquired (panel d). CL spectra of a pristine area (red line), thickness step (green line) and deformed crystal area (blue line). F), g) and h) are the monoCL maps acquired at 5.75 eV, 3.88 eV and 2.16 eV, respectively.**